# From Flip FET to Flip 3D Integration (F3D): Maximizing the Scaling Potential of Wafer Both Sides Beyond Conventional 3D Integration

Heng Wu†, Haoran Lu+, Wanyue Peng+, Ziqiao Xu, Yanbang Chu, Jiacheng Sun, Falong Zhou, Jack Wu, Lijie Zhang, Weihai Bu, Jin Kang, Ming Li, Yibo Lin, Runsheng Wang, Xin Zhang, Ru Huang

School of Integrated Circuits, Peking University, Beijing 100871, China, †email: hengwu@pku.edu.cn

+These authors contribute equally

**Abstract**

In this work, we proposed a new 3D integration technology: the Flip 3D integration (F3D), consisting of the 3D transistor stacking, the 3D dual-sided interconnects, the 3D die-to-die stacking and the dual-sided Monolithic 3D (M3D). Based on a 32-bit FFET RISCV core, besides the scaling benefits of the Flip FET (FFET), the dual-sided signal routing shows even more routing flexibility with 6.8% area reduction and 5.9% EDP improvement. Novel concepts of Multi-Flipping processes (Double Flips and Triple Flips) were proposed to relax the thermal budget constraints in the F3D and thus support the dual-sided M3D in the F3D. The core's EDP and frequency are improved by up to 3.2% and 2.3% respectively, after BEOL optimizations based on the Triple Flips compared with unoptimized ones.

## Introduction

As the conventional Moore's Law coming to an end, the semiconductor industry is thriving for new methods to continuously enhance the chip integration density, mainly following two technical routes. One is pushing the scaling of advanced logic technology by DTCO, trending towards 3D transistor stacking such as the Complimentary FET (CFET) [1-3] and backside (BS) interconnects [4-9]. We have proposed the FFET previously [10] as a combination of 3D integration of 3D stacked channel FETs and dual-sided interconnects, acting as a great candidate in the 2D & 3D transistor integration roadmap as illustrated in Fig. 1. The other one is the 3D IC including the 3D die stacking enabled by the 3D packaging [4,11] or the 3D tier stacking such as the Monolithic 3D integration (M3D) [12-14].

In this work, we proposed a new 3D integration technology called the ***Flip 3D*** integration (F3D) named after the repeated usage of wafer flipping process as an extension of the FFET. The F3D is the **first** to unite the 3D transistor stacking, 3D dual-sided interconnects, 3D face-to-face/back-to-back/face-to-back die stacking and the dual-sided M3D. For below, we will first review the roadmap of dual-sided interconnects and show the block-level power-performance-area (PPA) benefits, taking FFET as an example. Then we will discuss the F3D's unique Multi-Flipping processes, with much relaxed thermal budget constraint. At last, the future blueprint of F3D will be given, validating the great potentials combining the benefits of existing 3D integration technologies for the future.

## Dual-sided Interconnects

Dual-sided Interconnects (DSI) originates from the BS power delivery network (BSPDN) [4-8]. BS metals with much larger pitches are suitable not only for the power to reduce IR drop [4-7], but also for delivering timing-sensitive signals (such as the clock) to reduce the delay [7]. In ***DSI 1.0***, std. cells only have pins on the FS. Therefore, backside signal routing should be implemented by the Signal Transfer Cell (STC) [8,9] or the nTSV [7] which transfer signals to the other side but with area penalties, which is shown in Fig. 2 (a). Considering the ultra-scaled std. cells, routing resources on the frontside (FS) become increasingly insufficient. Thus, the std. cell design needs not only BS power but also BS signals [2,3,8,13]. In ***DSI 1.5***, std. cells are designed with BS intra-cell routing, so pins are naturally located on both FS and BS. However, in DSI 1.5, when the output pin of the current stage and the input pin of the next stage are on different sides (Fig. 2(b)), STCs or nTSVs are required to connect them. This leads to huge area loss because this kind of FS-BS interconnection occurs frequently in DSI 1.5. To solve this, in ***DSI 2.0*** we proposed the dual-sided output pin [10] existing in ***both FS and BS*** in the FFET which is composed of the common drain of nFETs and pFETs enabled by the Drain Merge structure given in Fig. 2(c). Thus, the FS signal routing and BS signal routing can be implemented independently, free of the usage of STCs or nTSVs. This results in significant area benefits w.r.t DSI 1.5. Moreover, STCs can still be added in P&R flow selectively to reduce redundant net length as illustrated in Fig. 2(c). Benchmark of these three DSI are listed in Table 1.

Physical implementation with DSI 2.0 and PPA analysis were carried out on a FFET 32-bit RISCV core (no cache, only computing core) based on the design rules listed in Table 2. Fig. 3 shows the post-P&R core layouts comparing the FS-only signal routing in FFET with $FP_1BP_0$ (100% FS pins and 0% BS pin: all pins set on the FS) & $FM_{0\text{-}8}BM_{0\text{-}1}$ (FM1-FM8 for inter-cell routing, M0 & M1 for intra-cell routing, etc.) and the DSI 2.0 in FFET with $FP_{0.5}BP_{0.5}$ (50% FS pins and 50% BS pins: input pins distributed evenly on both sides) & $FM_{0\text{-}4}BM_{0\text{-}4}$ at respective maximum utilization, the latter giving the area reduction of 6.8% thanks to the better routability by using the DSI 2.0. Fig. 4 shows the BS metal layer number can be reduced to 3 (even with BM1 & BM2 pitches enlarged by ×1.5) or even 2 by redistributing the input pin density, with some EDP degradation when the BS metal pitch increases or the BS metal layer number reduces. Thus, this trade-off between manufacturability and design space in the FFET should be carefully balanced. The successful implementation of the DSI 2.0 paves the way to set the I/O bumps on both sides, enabling the data communication between the face-to-face/back-to-back/face-to-back stacked dies in the F3D on both sides.

**Multi-Flipping Processes**

In the FFET process flow proposed previously [10], the FS RMG is formed before the BS epitaxy with single use of the wafer flipping process (***Single Flip***) for conceptual demonstration purpose. However, it is not compatible with the gate-last process in the advanced logic nodes. Though FS gate-first devices or low temperature BS epitaxy [15] can relax the thermal budget constraint, the device performance and reliability would degrade inevitably. To solve this, we proposed the ***Double Flips*** process, as illustrated in Fig, 5(a). Different from the Single Flip, the first wafer flipping is done right after the FS S/D epitaxy formation in the Double Flips. After the BS processes, the second wafer flipping is conducted, followed by the FS RMG process. Thus, both FS and BS metal gates are formed after the high temperature epitaxy. However, the BS MOL & BEOL processes in Double Flips still suffers from the high temperature FS RMG formation which is done after them. Metals and dielectrics with better high temperature stability (such as W with higher ρ and $SiO_2$ with higher κ) are required. This can be solved by adding another wafer flipping process (***Triple Flips***) in which both FS and BS MOL & BEOL are formed after all FEOL process, as given in Fig. 5(b). Detailed benchmarks of these 3 types of processes are listed in Table 3.

To investigate the performance difference between the Double and Triple Flips, 3 types of Double Flips (I-III) and 2 types of Triple Flips (IV-V) were studied at the block level, considering different BEOL metal and dielectric combos, which are shown in Table 4, taking Double Flips I as the baseline. Note that we assumed the same intrinsic device performance in both the Double & Triple Flips. For the Double Flips, the EDP slightly decreases by 0.2% and 0.5% (Fig. 6 (a)) and the frequency slightly increases by 0.1% and 0.7% (Fig. 6(b)) after reducing the BS metal resistance by replacing W with dual damascene Ru and subtractive Ru [16], respectively. In the Triple Flips IV, performance improves distinctly with the EDP decreases by 2.0% and the frequency increases by 1.1%. These metrics moves up to 3.2% and 2.3% respectively in the Triple Flips V for the lower via resistivity in the subtractive Ru interconnects.

101-stage ROs with fan-out 3 and distributed FS and BS BEOL loads were also simulated (Fig. 7(a)). The BEOL net length and via number are extracted from the P&R critical path statistics [17] of the FFET for each side separately, as shown in Fig. 7(b-c) (BS data are not shown). Taking Double Flips I as the baseline, the RO frequency in the Triple Flips V increases by 7.04% and 9.91% @ Vdd = 0.7 V with 1 um and 3 um BEOL loads, respectively.

By using Multi-Flipping processes, the F3D features more manufacturing-friendly runpath with clear performance gains.

**Flip 3D Integration**

The nowadays advanced integrated circuits mainly focus on two topics. One is the advanced CMOS, trending towards 3D integration of 3D FETs and 3D interconnects, as given in Fig. 8(a). The CFET, though a possible solution to 3D transistor stacking, faces great challenges of high aspect ratio (AR) process. The other one is the 3D IC, enabled by advanced packaging or the M3D, as depicted in Fig. 8(b). However, the advanced packaging is limited by the single-sided hybrid bonding processes to enable further stacking due to the lack of I/O bumps on the other side of dies. While, the M3D technology only explores the FS usage of the wafer.

The F3D technology could act as the ultimate form of these 3D integration technologies on the roadmap (Fig. 8(c)) as following. Firstly, the FFET (demonstrated in Die I) is more manufacturing-friendly for realizing the 3D integration of 3D FETs than the CFET because of the lower AR processes [10] thanks to the independent process on each side of the wafer. Secondly, the F3D supports the dual-sided I/O bumps thanks to the DSI 2.0. Thus, both FS and BS of chip dies support the hybrid bonding, enabling the free choice of face-to-face/back-to-back/face-to-back bonding. It could further remove the TSVs used in the 3D packaging applications such as the HBM [11] (demonstrated in Die II). Thirdly, with the help of the Multi-Flipping processes, the F3D also supports the M3D on both FS and BS of the wafer (demonstrated in Die III) with much relaxed thermal budget constraints.

Overall, the F3D fully utilizes both sides of the wafer, pushing the logic/memory density and their co-integration to the limits, and extends the boundary of the 3D packaging, enabling even broader chip stacking space for future.

**Conclusion**

With the manufacturing-friendly 3D stacked transistor architecture (such as FFET) supported by Multi-Flipping processes, the 3D interconnects enabled by the DSI 2.0 and the dual-sided 3D die stacking enabled by the dual-sided hybrid bonding, the F3D shows great potential of opening up new era for the next-generation 3D integration.

## Acknowledgments

This work was supported by the National Key R&D Program of China under Grant 2023YFB4402200; and in part by the 111 Project under Grant 8201702520.

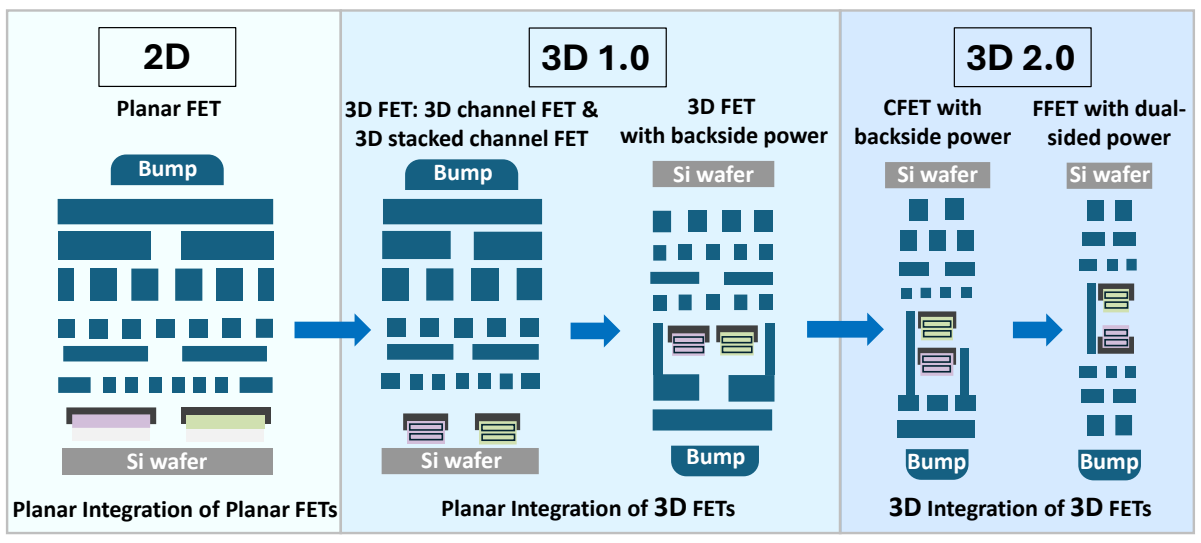

**Fig. 1** The roadmap of 2D & 3D transistor integration. The 3D transistor integration began with the 3D FET (***3D 1.0***) and is now trending towards BS interconnects to achieve further scaling. The ***3D 2.0*** stands for 3D integration of 3D FETs including 3D stacked FETs and dual-sided interconnects. The FFET [10] acts as a great candidate for 3D transistor stacking technology beyond the CFET [1-3].

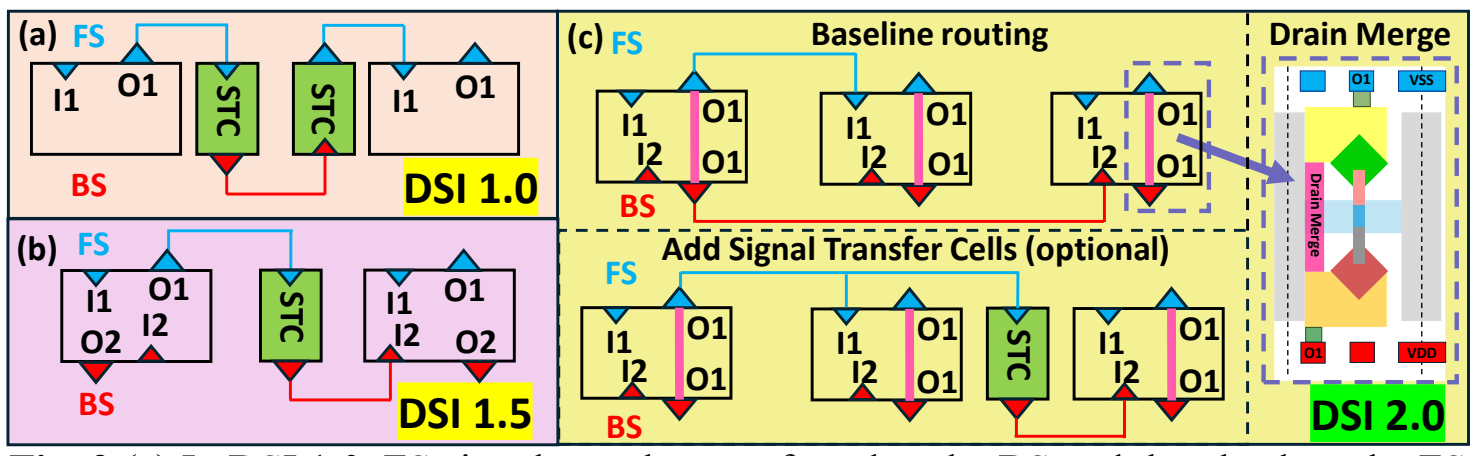

**Fig. 2** (a) In DSI 1.0, FS signals can be transferred to the BS and then back to the FS through the STCs. (b) In DSI 1.5, substantial STCs should be inserted to connect the pins on different sides. (c) In DSI 2.0, the dual-sided output pin enables independent dual-sided routing, making the STC an optimizing but not necessary option.

| | DSI 1.0 [7,9] | DSI 1.5 [2,3,8,13] | DSI 2.0 [10] |
|---|---|---|---|
| Std. cell pin location | FS | FS and BS | FS and BS |
| Output pin category | FS | Either FS or BS | Both FS and BS |
| BS intra-cell routing | No | Yes | Yes |
| nTSV / Signal Transfer Cell | Necessary | Necessary | Optional |
| Area waste due to inter-side routing | Moderate | Large | Small |

**Table 1** Benchmarks of the dual-sided interconnects of the three gen0erations.

| Layers | Pitch |
|---|---|
| FM8-FM5 | 76 nm |
| FM4-FM3 | 42 nm |
| FM2 | 30 nm |
| FM1 | 34 nm |
| FM0 | 28 nm |
| Poly | 51 nm |
| BM0 | 28 nm |
| BM1 | 34 nm |
| BM2 | 30 nm |
| BM3-BM4 | 42 nm |
| BM5-BM6 | 76 nm |

**Table 2** FEOL and BEOL design rules used in this work.

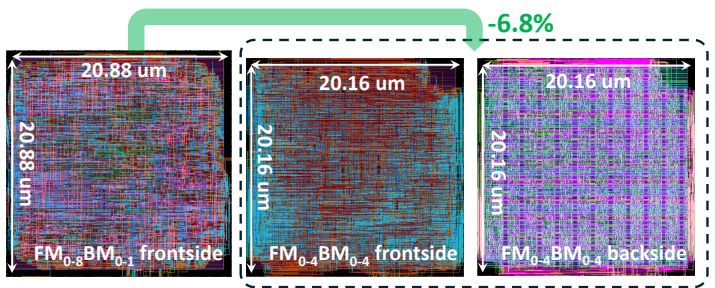

**Fig. 3** Post-P&R minimum core layouts comparing the FS-only signal routing in FFET $FP_1BP_0$ & $FM_{0\text{-}8}BM_{0\text{-}1}$ and the DSI 2.0 in the FFET $FP_{0.5}BP_{0.5}$ & $FM_{0\text{-}4}BM_{0\text{-}4}$. The DSI 2.0 gained area reduction of 6.8% over the FS-only signal routing.

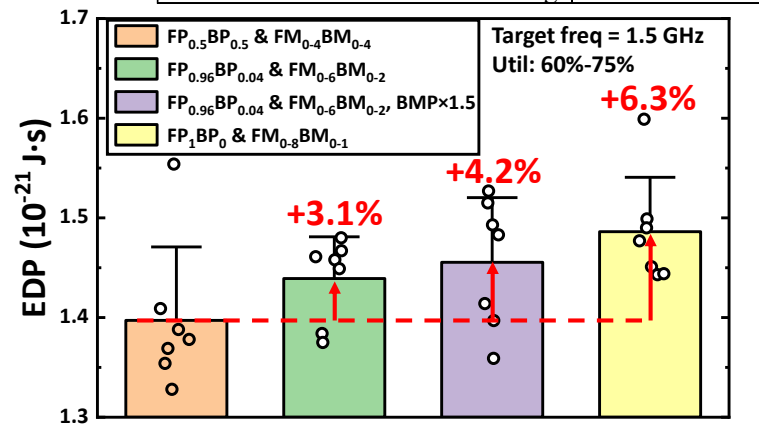

**Fig. 4** The block's EDP degrades as the BS metal layer number reduces to 2 by allocating BS input pins to the FS. BS metal pitch could be enlarged with slight EDP degradation.

| | Single Flip | Double Flips | Triple Flips |
|---|---|---|---|
| Thermal budget issues | FS Gate, FS BEOL | BS BEOL | None |
| Cost | Low | Medium | High |
| FEOL choices | FS gate-first devices / BS epitaxy with low thermal budget [15] | No limitation | No limitation |
| BEOL choices | FS: W/Ru + Higher κ dielectric BS: No limitation | FS: No limitation BS: W/Ru + Higher κ dielectric | No limitation |

**Table 3** Benchmarks of Multi-Flipping processes.

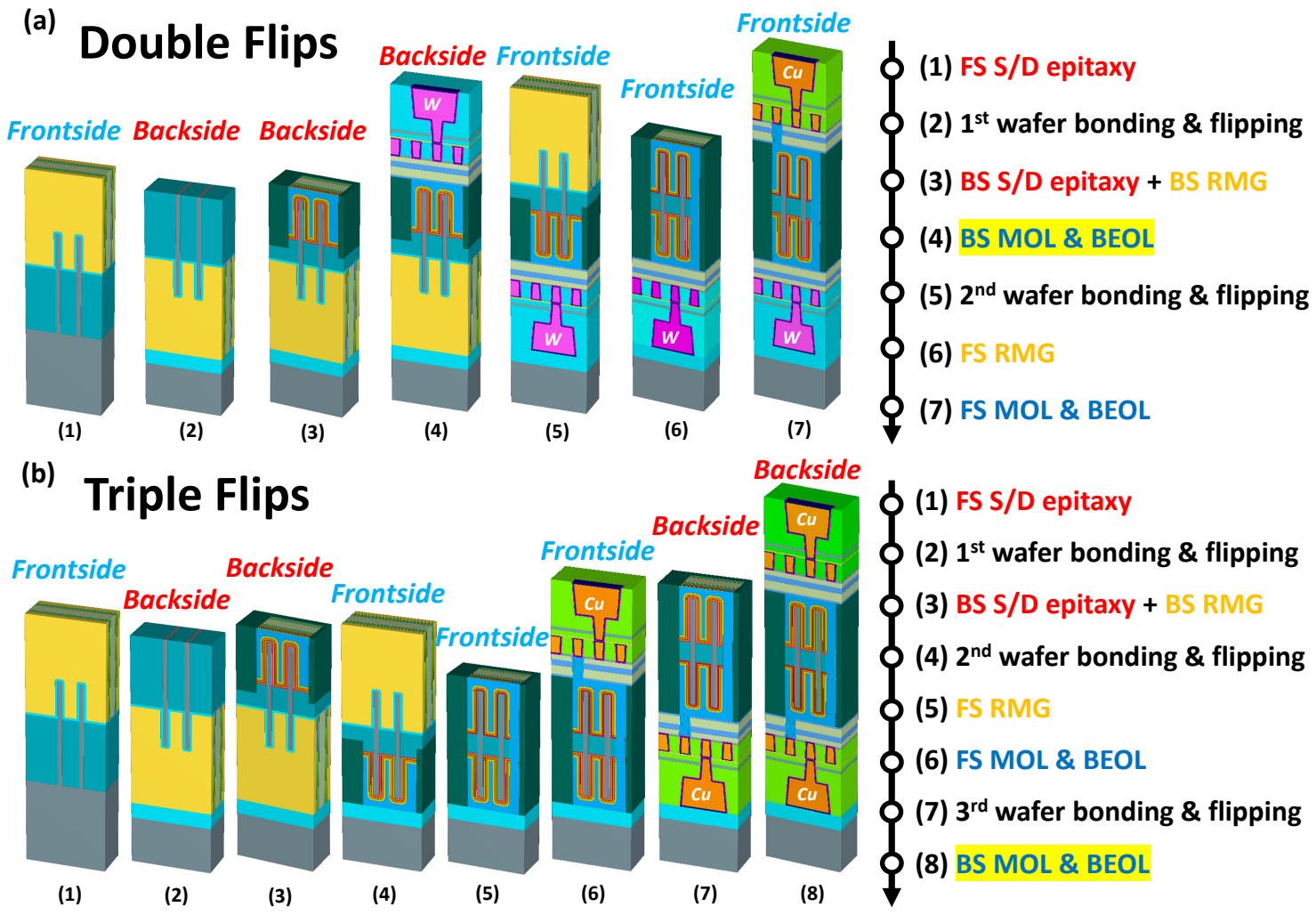

**Fig. 5** (a) Process flow of the Double Flips. FS and BS metal gates are both formed after the high temperature epitaxy. (b) Process flow of the Triple Flips. The BS MOL & BEOL are formed at last by flipping the wafer right after the BS RMG in step (3).

| | Flip Type | FS BEOL Metal | FS BEOL Dielectric | BS BEOL Metal | BS BEOL Dielectric |
|---|---|---|---|---|---|
| I | Double | Cu | κ = 3 | W | κ = 4 |
| II | Double | Cu | κ = 3 | dual damascene Ru | κ = 4 |
| III | Double | Cu | κ = 3 | subtractive Ru | κ = 4 |
| IV | Triple | Cu | κ = 3 | Cu | κ = 3 |
| V | Triple | subtractive Ru | κ = 3 | subtractive Ru | κ = 3 |

**Table 4** 3 types of Double Flips and 2 types of Triple Flips with different process assumptions. Metal resistivity is based on [18,19]. The subtractive Ru has lower via resistance than the dual damascene Ru due to the subtractive top via with reduced liner thickness [16].

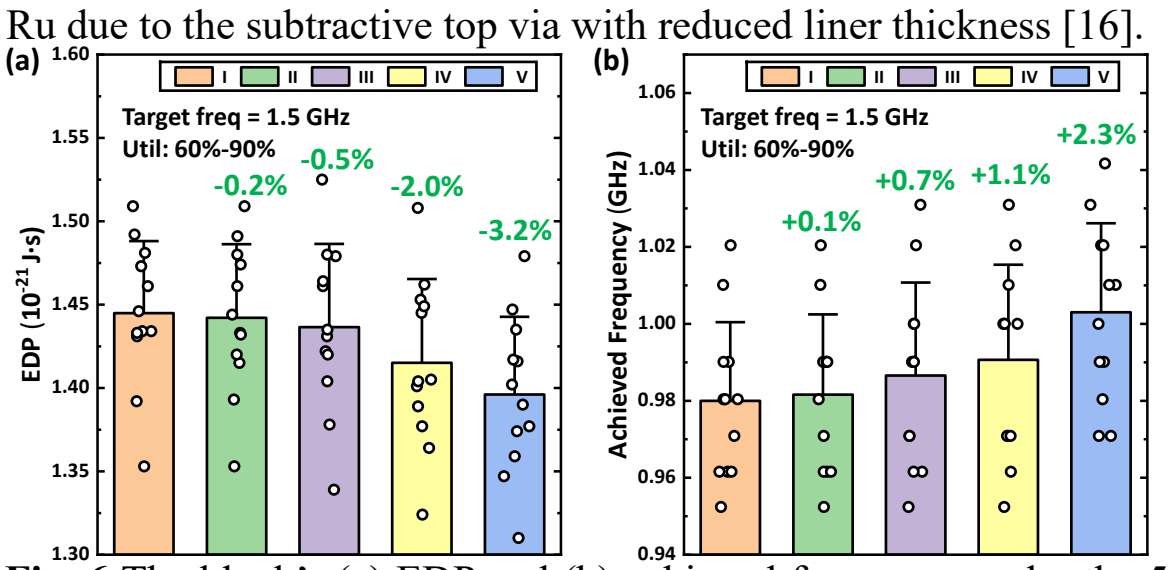

**Fig. 6** The block's (a) EDP and (b) achieved frequency under the 5 types of experiments listed in Table 4. All experiments used DSI 2.0 in FFET $FP_{0.5}BP_{0.5}$ $FM_{0\text{-}4}BM_{0\text{-}4}$. The FSPDN for the Double Flips and the BSPDN for Triple Flips were removed for the fair comparison.

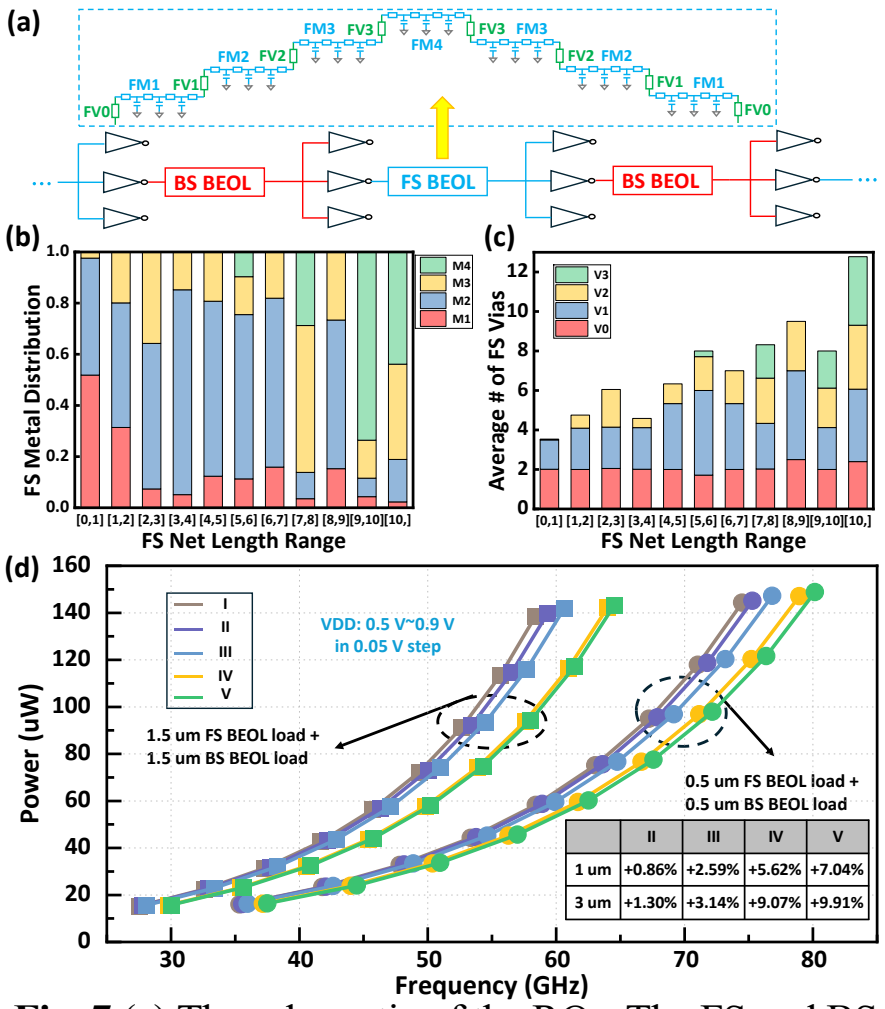

**Fig. 7** (a) The schematic of the ROs. The FS and BS nets are set alternately to imitate the nets in FFET $FP_{0.5}BP_{0.5}$ & $FM_{0\text{-}4}BM_{0\text{-}4}$. (b) FS net length statistics. (c) FS via statistics. (d) Power-freq plot with 2 BEOL loads. Table inset is the frequency gain w.r.t case I.

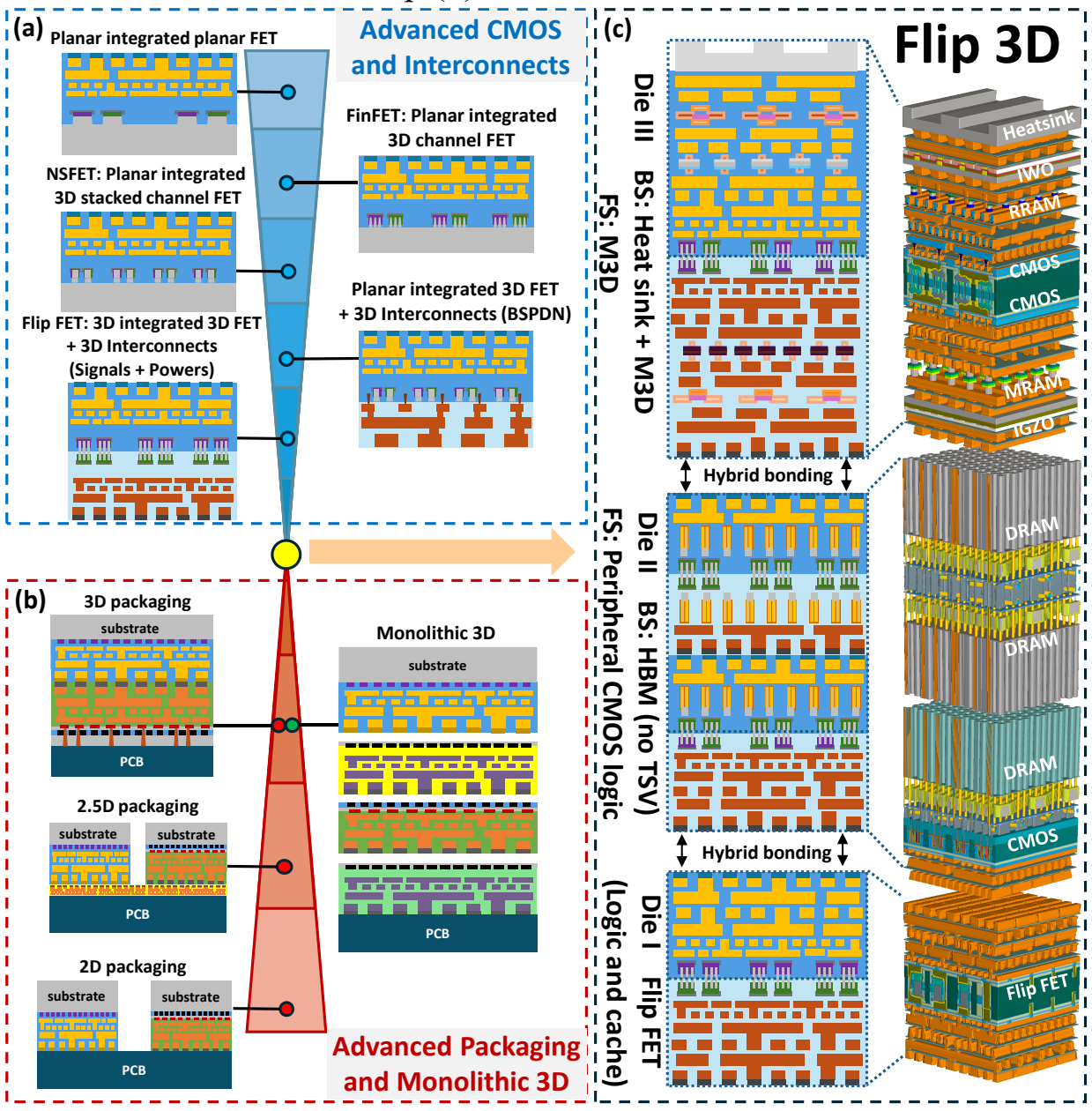

**Fig. 8** (a) The roadmap of the advanced CMOS and interconnects trending towards the 3D transistor integration and 3D interconnects integration. (b) The roadmap of the advanced packaging and the Monolithic 3D which are both approaches to the 3D IC. (c) 2D schematic and 3D detailed structure demo of the F3D. The three stacked dies are fabricated separately by Multi-Flipping processes before being face-to-face/back-to-back/face-to-back stacked on each other by the hybrid bonding. The F3D is the combination of the 3D transistor stacking (Die I), the 3D die stacking (Die II) and the dual-sided M3D (Die III) and the DSI on the FS and BS of all the dies.